\documentclass[prb,superscriptaddress,twocolumn,showpacs]{revtex4}
\usepackage{epsfig}
\begin{document}
\title{Stability of negative and positive trions in quantum wires}

\author{B. Szafran}
\affiliation{Departement Fysica, Universiteit Antwerpen (Campus
Drie Eiken), Universiteitsplein 1, B-2610 Antwerpen, Belgium}
\affiliation{Faculty of Physics and Applied Computer Science, AGH
University of Science and Technology, al. Mickiewicza 30, 30-059
Krak\'ow, Poland}
\author{T. Chwiej}
\affiliation{Faculty of Physics and Applied Computer Science, AGH
University of Science and Technology, al. Mickiewicza 30, 30-059
Krak\'ow, Poland}
\author{F.M. Peeters}
\affiliation{Departement Fysica, Universiteit Antwerpen (Campus
Drie Eiken), Universiteitsplein 1, B-2610 Antwerpen, Belgium}
\author{S. Bednarek}
\affiliation{Faculty of Physics and Applied Computer Science, AGH
University of Science and Technology, al. Mickiewicza 30, 30-059
Krak\'ow, Poland}
\author{J. Adamowski}
\affiliation{Faculty of Physics and Applied Computer Science, AGH
University of Science and Technology, al. Mickiewicza 30, 30-059
Krak\'ow, Poland}
\date{\today}

\begin{abstract}
Binding energies of negative ($X^-$) and positive trions ($X^+$)
in quantum wires are studied for strong quantum confinement of
carriers which results in a numerical exactly solvable model. The
relative electron and hole localization has a strong effect on the
stability of trions. For equal hole and electron confinement,
$X^+$ is more stable but a small imbalance of the particle
localization towards a stronger hole localization e.g. due to its
larger effective mass, leads to the interchange of $X^-$ and $X^+$
recombination lines in the photoluminescent spectrum as was
recently observed experimentally. In case of larger $X^-$
stability, a magnetic field oriented parallel to the wire axis
leads to a stronger increase of the $X^+$ binding energy resulting
in a crossing of the $X^+$ and $X^-$ lines.
\end{abstract}
 \maketitle
\section{Introduction}
Trions are charged exciton complexes formed when an electron or a
hole is bound\cite{Lampert} to a neutral exciton ($X$). The
binding energies of the complexes are very small in bulk, but they
are substantially enhanced in structures of reduced
dimensionality, i.e., in quantum
wells\cite{Stebe2D,Riva2D,Moradi,Hawrylak,Whittaker,Chapman,Glasberg}
and quantum wires.\cite{Esser,MCTs,Otterburg}

Due to the larger effective mass of the hole, in bulk\cite{SM} as
well as in strictly two-dimensional confinement\cite{Riva2D} the
binding energy of positive trions ($X^+$) is larger than the
negative trion ($X^-$) binding energy.  However, in quantum wells
the observed\cite{Glasberg} $X^-$ and $X^+$ binding energies are
nearly equal, which is explained\cite{Glasberg,Moradi} by a
stronger hole localization within the quantum well enhancing the
hole-hole interaction. The magnetic field perpendicular to the
plane of confinement enhances more strongly the $X^-$ stability
leading to a crossing of $X^-$ and $X^+$ binding
energies.\cite{RV,yusa} For trions localized on a defect of the
quantum well potential $X^-$ can become more stable than $X^+$
even without the presence of an external magnetic
field.\cite{ClaraMC} The combined quantum well and defect
confinement creates a three-dimensional potential similar to a
quantum dot. In quantum dots the localization-related hole-hole
interaction enhancement leads to the interchange of the order of
the $X^-$ and $X^+$ recombination lines in the photoluminescence
(PL) spectrum already for quantum dot diameters as large as 24
donor Bohr radii.\cite{Szafran1} For smaller dots the $X^+$ line
becomes even more energetic\cite{Szafran1,Szafran2} than the $X$
line. In coupled dots\cite{Szafran2,Belluci} this effect leads to
the ground-state dissociation of $X^+$,\cite{Szafran2,Belluci} for
which the holes in the ground state occupy different dots.

The present work is motivated by a recent experimental
study\cite{Otterburg} of positive and negative exciton trions in
V-groove GaAs/AlGaAs quantum wires. The negative trion was found
to be distinctly more stable than $X^+$ (binding energies of $X^-$
and $X^+$ were determined as 4.2 and 2.9 meV, respectively). Here,
we indicate that the observed\cite{Otterburg} order of $X^-$ and
$X^+$ energy lines may be a consequence of modifications of the
interactions due to a stronger hole confinement. In a previous
theoretical study\cite{Esser} of trions in quantum wires $X^+$ was
found to be more stable than $X^-$, which was obtained in the case
of equal hole and electron confinement. A crossing of $X^-$ and
$X^+$ PL lines as functions of the wire width has previously been
obtained in a quantum Monte-Carlo study\cite{MCTs} of a quantum
wire with a square well confinement potential. In this paper we
focus on the effect due to different electron and hole
localization leading to modifications of the effective
inter-particle interactions. We study the correlations between
electrons and holes and consider the effect of a magnetic field
oriented parallel to the quantum wire. The study of the stability
of the trions is performed as function of the electron and hole
localization instead of dimensions of the wire. It has been
demonstrated\cite{Molinari} that in realistic quantum wires with
strong confinement the binding energy of neutral excitons is
governed by a size dependent parameter independent of the shape
and composition of the wire.

The stronger hole localization results from its weak penetration
into the barrier material due to its larger effective mass than
the electron band mass. For the spillover of the electron wave
function out of the quantum wire, recently observed in
self-assembled InAs/InP quantum wires,\cite{Yosip} the ratio of
the electron to hole localization can in principle be arbitrarily
large.\cite{Yosip} However, in the following we show that even a
{\it small} enhancement of the hole localization changes the order
of the $X^-$ and $X^+$ PL recombination lines.

For the purpose of the present study we apply the single band
model for the hole and consider a harmonic oscillator confinement
potential in the directions perpendicular to the wire, referred to
as "lateral" in the following. The present model does not account
for the interface between the wire and barrier materials, so the
effective mass discontinuity and dielectric constant mismatch are
neglected. These effects usually strengthen the electron-hole
interaction and weaken the penetration of the wave functions into
the barrier. They are however of a secondary importance for
GaAs/AlGaAs,\cite{GaAs} InAs/InP,\cite{InAsInP} and
CdTe/ZnTe\cite{CdTe} quantum wires. Note that, the present
modelling is inapplicable to the free-standing quantum wires,
where the image charge effect is extremely strong.\cite{Xia}

We assume that the lateral confinement is strong, so that only the
lowest subband for the electron and hole is occupied. This
assumption allows for a reduction of the Schr\"odinger equation to
an effective two-dimensional form. Usually the solution of the
trion eigenequations is very challenging and requires extensive
variational
calculations\cite{Stebe2D,Riva2D,Moradi,Hawrylak,Whittaker,SM,Szafran1,Szafran2}
or application of the quantum Monte Carlo
methods.\cite{ClaraMC,MCTs} The present problem is unique in the
sense that it allows for an exact inclusion of the interparticle
correlations.

The paper is organized as follows: the next Section contains the
theory, the results are given in Section III, the conclusion and
summary are presented in Section IV.

\section{Theory}

We adopt the donor units, i.e., donor Bohr radius
$a_d=4\pi\epsilon_0\epsilon\hbar^2/m_ee^2$ for the unit of length
and twice the donor Rydberg $2R_d=\hbar^2/m_ea_d^2$ as the unit of
the energy, where $m_e$ is the band electron effective mass and
$\epsilon$ is the dielectric constant. In these units, the
Hamiltonian for a single electron in a quantum wire with harmonic
oscillator lateral confinement has the form
\begin{equation}
H_e=-\frac{1}{2}\frac{\partial^2}{\partial z_e^2}+H^l_e,
\label{seh}
\end{equation}
with the lateral Hamiltonian \begin{equation}
H^l_e=-\frac{1}{2}\left(\frac{\partial^2}{\partial x_e^2}+
\frac{\partial^2}{\partial
y_e^2}\right)+\frac{1}{2l_e^4}(x_e^2+y_e^2),
\end{equation}
where $l_e$ is the length of the harmonic oscillator confinement
for the electron. The ground-state wave function of the
Hamiltonian (2) is $\Psi_e=\exp[-(x^2+y^2)/2l_e^2]/l_e\sqrt{\pi}$
with the energy eigenvalue $E_e=1/l_e^2$. In the adopted
single-band approximation the hole ground-state wave function
($\Psi_h$) of the lateral confinement has the form of $\Psi_e$ but
with $l_h$ - the harmonic oscillator length for the hole instead
of $l_e$, and the energy is $E_h=1/\sigma l_h^2$, where
$\sigma=m_h/m_e$ is the hole to electron effective mass ratio, or,
in other words, the hole mass in the donor units. The negative
trion Hamiltonian can be written as
\begin{equation}
H_-=H_{e1}+H_{e2}+H_h-\frac{1}{r_{e1h}}-\frac{1}{r_{e2h}}+\frac{1}{r_{12}},
\end{equation}
where $r_{e1h}$ ($r_{e2h}$) is the distance between the first
(second) electron and the hole and $r_{12}$ is the
electron-electron distance. We assume that the lateral confinement
is sufficiently large that the trion wave function can be
effectively separated into a product \begin{eqnarray} &\psi({\bf
r}_{e1},{\bf r}_{e2},{\bf r}_{h})= &
\Psi(x_{e1},y_{e1})\Psi(x_{e2},y_{e2}) \Psi(x_{h},y_{h}) \nonumber
\\ & & \times \chi_-(z_{e1},z_{e2},z_h), \label{ps}
\end{eqnarray}
where $\chi_-$ is the negative trion wave function of the motion
along the wire. The Hamiltonian (3) integrated over the lateral
degrees of freedom with the wave function (\ref{ps}) produces the
effective trion Hamiltonian:
\begin{eqnarray}
&H_-^\mathrm{ef}=&-\frac{1}{2}\left(\frac{\partial^2}{\partial
z_{e1}^2}+\frac{\partial^2}{\partial
z_{e2}^2}\right)-\frac{1}{2\sigma}\frac{\partial^2}{\partial
z_h^2} \nonumber
\\ & & +V^\mathrm{ef}(l_e;z_{e1}-z_{e2})
-V^\mathrm{ef}(l_{eh};z_{e1}-z_{h})\nonumber \\ & &
-V^\mathrm{ef}(l_{eh};z_{e2}-z_{h}), \label{kuua}
\end{eqnarray}
with $l_{eh}=\sqrt{(l_e^2+l_h^2)/2}$ and the effective interaction
potential\cite{Esser,1D}
\begin{equation}
V^\mathrm{ef}(l;z)=(\pi/2)^{1/2}\mathrm{erfc}(|z|/\sqrt{2}l)\exp(z^2/2l^2)/l,
\end{equation}
which is finite at the origin ($V^\mathrm{ef}(l;0)=1/l$) and
approaches the $1/z$ asymptotic at large $z$. Hamiltonian (5) is
written with respect to the sum of the ground-state energies of
noninteracting two electrons and one hole. Therefore, the absolute
value of the (negative) energy of a bound state is interpreted as
the energy needed to separate all the particles away from one
another. Introducing the center-of-mass coordinate
$Z=\left(z_{e1}+z_{e2}+\sigma z_h\right)/(2+\sigma)$ one obtains
$H_-^\mathrm{ef}=-\frac{1}{2M}\frac{\partial^2}{\partial
Z^2}+H_-^\mathrm{rel}$, where $M=2+\sigma$ is the negative trion
mass and $H^\mathrm{rel}$ is the relative motion Hamiltonian
\begin{eqnarray}
&H_-^{\mathrm{rel}}&=-\frac{1}{2\mu}\left(\frac{\partial^2}{\partial
z_{h1}^2}+\frac{\partial^2}{\partial
z_{h2}^2}\right)-\frac{1}{\sigma}\frac{\partial^2}{\partial z_{h1}
\partial z_{h2}} \nonumber \\
& & +V^\mathrm{ef}(l_e;z_{h1}-z_{h2})-V^\mathrm{ef}(l_{eh};z_{h1})
\nonumber \\ & &  -V^\mathrm{ef}(l_{eh};z_{h2}), \label{ujo}
\end{eqnarray}
with the reduced mass of an electron-hole pair
$\mu=\sigma/(1+\sigma)$, and the coordinates of the relative
electron-hole positions $z_{h1}=z_h-z_{e1}$ and
$z_{h2}=z_h-z_{e2}$. In these coordinates the inter-electron
distance along the length of the wire is $z_{12}=|z_{h1}-z_{h2}|$.
The wave function $\chi_-$ is separable into a product of the
center of mass and relative wave function
$\chi_-(z_{e1},z_{e2},z_h)=\chi_{CM}(Z)\chi(z_{h1},z_{h2})$.

The corresponding relative Hamiltonian for the positive trion has
the following form:
\begin{eqnarray}
&H_+^{\mathrm{rel}}&=-\frac{1}{2\mu}\left(\frac{\partial^2}{\partial
z_{h1}^2}+\frac{\partial^2}{\partial
z_{h2}^2}\right)-\frac{\partial^2}{\partial z_{h1}
\partial z_{h2}} \nonumber \\
& & +V^\mathrm{ef}(l_h;z_{h1}-z_{h2})-V^\mathrm{ef}(l_{eh};z_{h1})
\nonumber \\ & &  -V^\mathrm{ef}(l_{eh};z_{h2}), \label{ujop}
\end{eqnarray}
with $z_{h1}$, $z_{h2}$ standing here for the relative position
coordinates of the first and second hole with respect to the
electron position. The reference energy for the Hamiltonian
(\ref{ujop}) is the energy of the dissociated complex, i.e,
$2E_h+E_e$.

In the following we consider also the exciton for which the
effective Hamiltonian written with respect to the energy of a
dissociated electron and hole pair reads
\begin{equation} H^X=-\frac{1}{2\mu}\frac{\partial^2}{\partial
z_{eh}^2}-V^\mathrm{ef}(l_{eh};z_{eh}). \label{HX}
\end{equation}
The lowest eigenvalue of this Hamiltonian is equal to minus the
exciton binding energy ($-E_B^{X}$). On the other hand the
difference between $-E_B^{X}$ and the eigenvalues of trion
Hamiltonians (\ref{ujo},\ref{ujop}) is equal to the trion binding
energies ($E_B^{X^-}$, $E_B^{X^+}$) with respect to dissociation
into an exciton and a free electron (for $X^-$) or a hole (for
$X^+$). Trion binding energies are equal to the red-shift of the
trion recombination lines with respect to the exciton line in the
PL spectrum.

\begin{figure}[htbp]
    \epsfxsize=58mm
                \epsfbox[98 450 372 716] {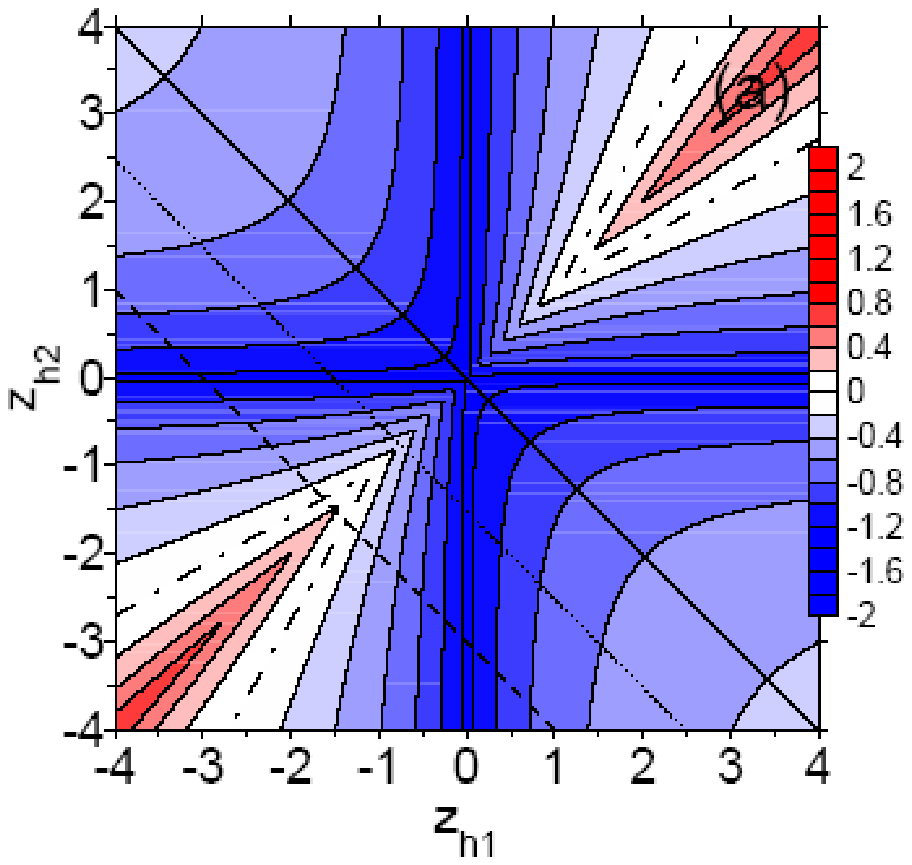}
    \epsfxsize=60mm
                \epsfbox[20 10 572 560] {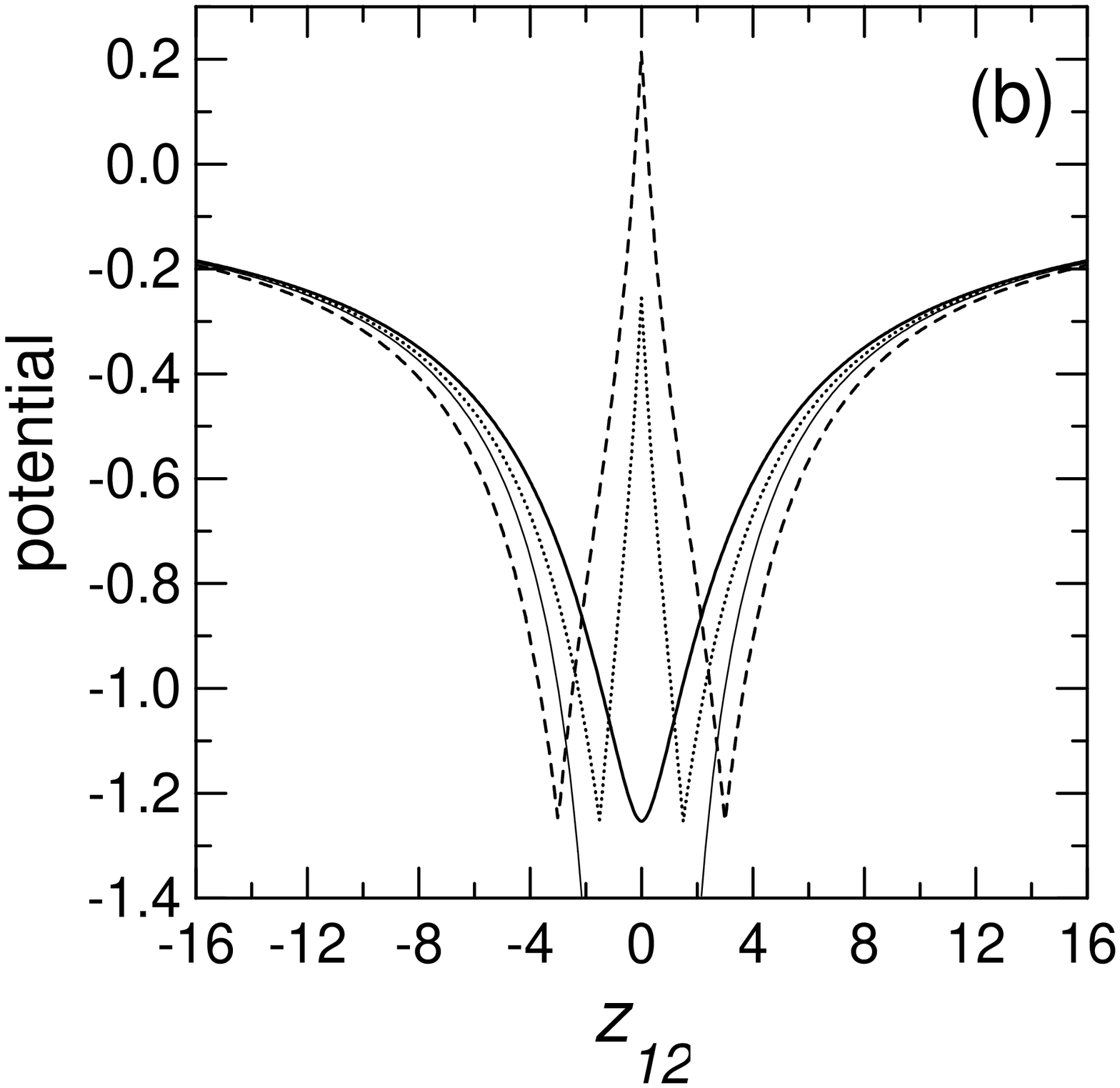}

\caption{(a) [color online] Contour plot of the interaction
potential
$V=V^\mathrm{ef}(L;z_{h1}-z_{h2})-V^\mathrm{ef}(L;z_{h1})-V^\mathrm{ef}(L;z_{h2})$
 as function
of the interparticle distances with lateral confinement length
$L=l_e=l_h=1$. Distances and energies are given in donor units.
The dashed-dotted line corresponds to $V=0$. (b) The interaction
potential plotted for $L=1$ along the lines $z_{h2}=-z_{h1}$,
$z_{h2}=-z_{h1}-1.5$, and $z_{h2}=-z_{h1}-3$ marked in (a) with
(thick) solid, dotted and dashed lines respectively, as function
of the interelectron ($X^-$) or interhole ($X^+$) distance
$z_{12}=z_{h1}-z_{h2}$. Thin solid line shows the $-3/|z_{12}|$
asymptotic. }
\end{figure}

We solve the relative Hamiltonian eigenequations using the
imaginary time technique\cite{ktd} on a two-dimensional grid with
a finite-difference approach. We use 201 points in both $z_{h1}$
and $z_{h2}$ directions. The size of the computational box in both
directions is chosen "self-consistently" to be 12 times larger
than the average distance between the particles of the same charge
defined as $<(z_{h1}-z_{h2})>^{1/2}$.

In the present calculations we assumed harmonic oscillator lateral
confinement which allows us to simplify the problem considerably
because of the availability of analytical formula\cite{Esser,1D}
for the effective one-dimensional interaction. The confinement
lengths $l_e$ and $l_h$ parametrize the strength of the particle
localization. Since the single-particle energies cancel in the
calculation of the trion binding energies, the applicability of
the present results is wider. In fact the present results can be
used for any form of the lateral confinement (which does not even
have to be cylindrically symmetric) as long as it produces the
same effective interaction potential. For instance the
electron-electron interaction potential for $l_e=2.95$ and 6 nm is
very well (i.e. with a precision better than $2\%$) reproduced for
a GaAs quantum wire ($m_e$=0.067) with a circular square well
confinement of depth 320 meV and diameters 9.6 nm and 22.8 nm
respectively. For elliptical harmonic oscillator confinement with
different oscillator lengths in $x$ and $y$ directions ($l_x$ and
$l_y$, respectively) we cannot give a closed analytical formula
for the effective interaction potential. Nevertheless, we have
found via a numerical integration that the interaction potential
between two electrons in an elliptical wire can be surprisingly
well reproduced by formula (6) for a circular wire with an
effective $l=(l_x+l_y)/2$. The numerically calculated deviation
between the two potentials is not larger than $2\%$ for any
interelectron distance. The essential assumption of the present
model therefore does not rely on the form of the lateral
confinement but on its strength, which has to be large enough to
prevent the Coulomb interactions from deforming the lateral wave
functions. The applied assumption of the frozen lateral degrees of
freedom for the electron and the hole is applicable for the
exciton binding energy when $l_e<a_d$ and $l_h<a_d$. This
condition guarantees that the length of the lateral confinement of
the carriers is smaller than the bulk exciton radius, and that the
sum of the lateral confinement energies for the electron and the
hole are at least two times larger than the exciton binding energy
in bulk. For trions the applied approximation is  better justified
and the conditions are less stringent because the trions have a
larger size and have smaller binding energies than the exciton.

\begin{figure*}[htbp]

    \epsfxsize=130mm
                \epsfbox[70 130 560 770] {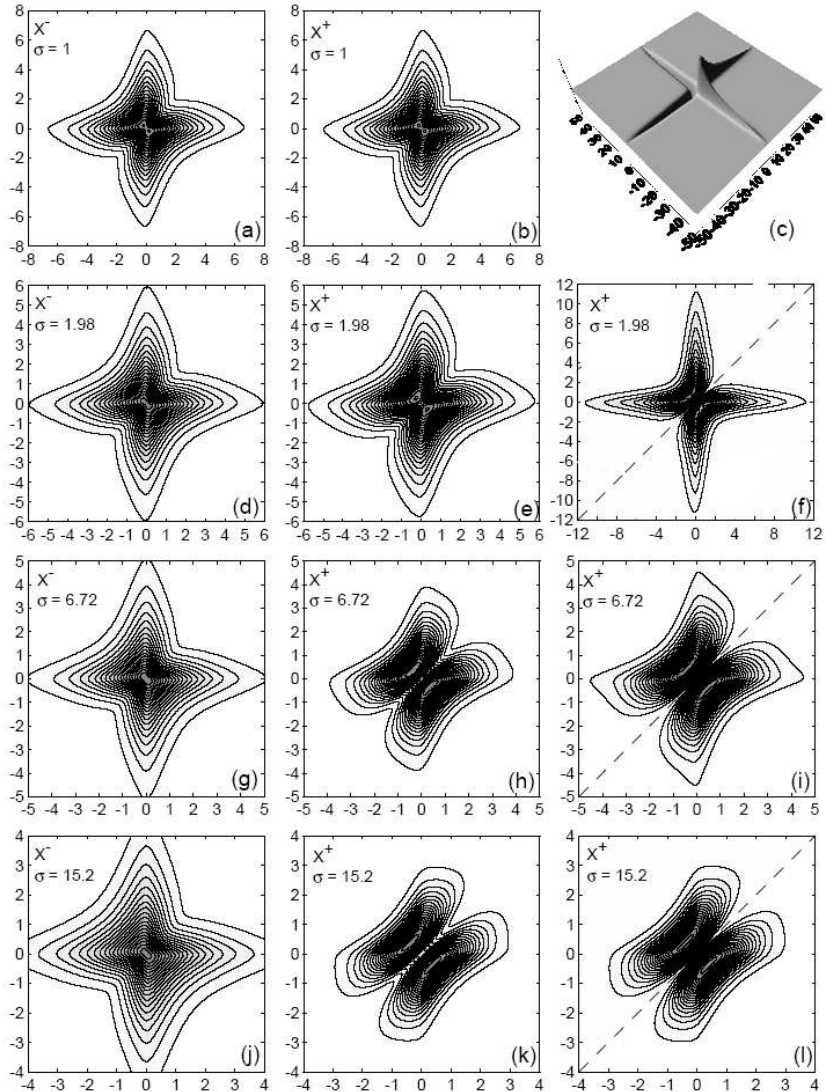}

\caption{Wave functions for negative (a,d,g,j) and positive
(b,c,e,f,h,i,k,l) trions for $l_e=l_h=L=0.2$ in $z_{h1}$ and
$z_{h2}$ coordinates (horizontal and vertical axis, respectively)
for different values of the mass ratio $\sigma$. Plots (c,f,i,l)
show the wave functions of the excited $X^+$ state antisymmetric
with respect to the interchange of the holes. The dashed line in
(f,i,l) shows the node of the wave functions. Plot (c) corresponds
to an unbound state, for other plots the computational box is
larger than the fragment displayed and the states are bound. }
\end{figure*}

\begin{figure}[hbtp]
    \epsfxsize=60mm
                \epsfbox[60 33 555 562] {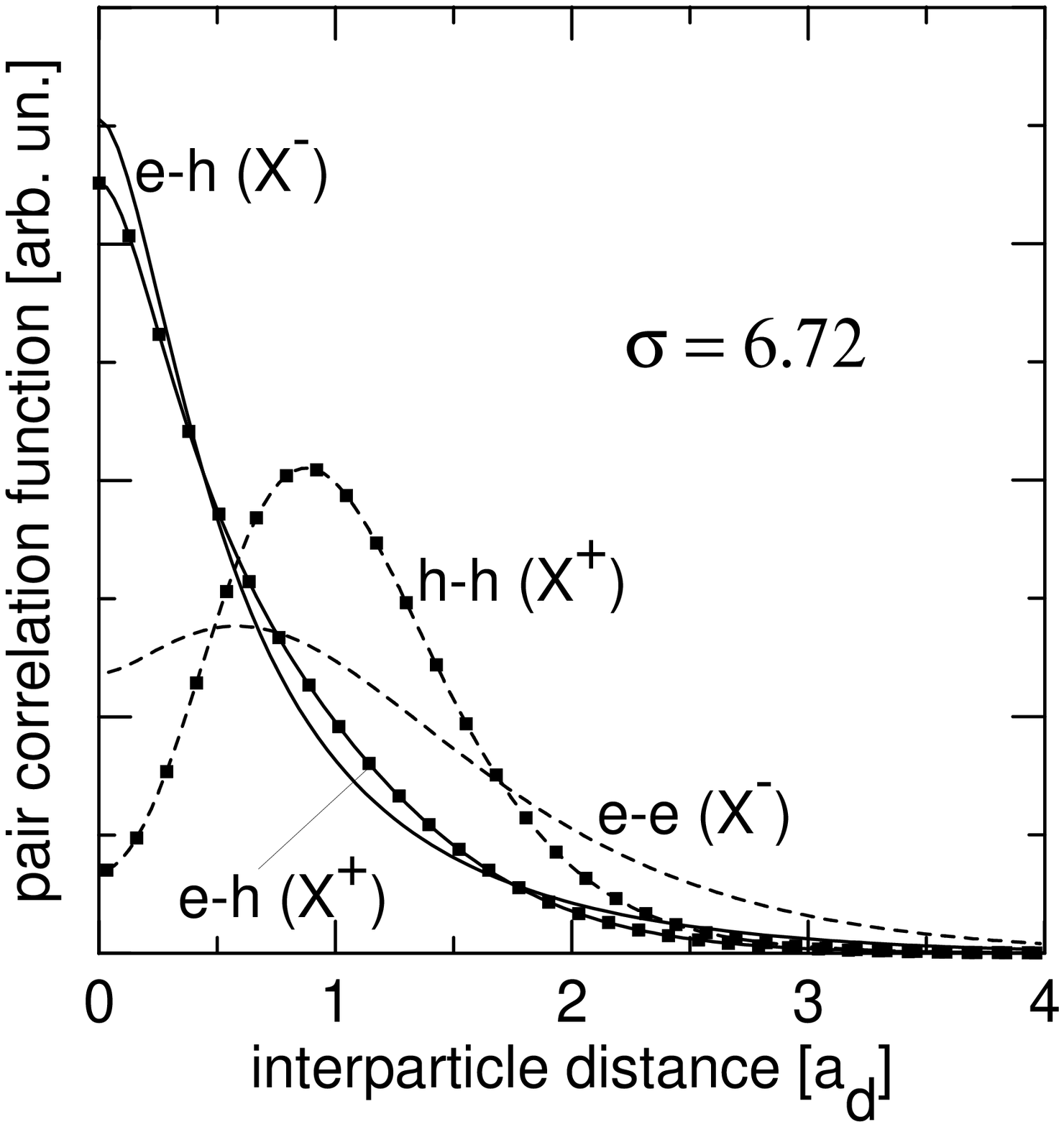}

\caption{Electron-hole (solid lines), electron-electron, hole-hole
(dashed lines) pair correlation functions plots for $X^-$ and
$X^+$ (lines marked by black squares) at $\sigma=6.72$ and
$l_e=l_h=0.2$ .}
\end{figure}

\section{Results}

We start the presentation of our results by discussing the
properties of trions in quantum wires with equal lateral
confinement for the electron and the hole (subsection III.A), and
then in subsection III.B we show the effect of different
confinements for the stability of $X^+$ and $X^-$ trions.
Subsection III.C describes the effect of the magnetic field
oriented parallel to the axis of the wire.

\begin{figure}[htbp]
    \epsfxsize=70mm
                \epsfbox[28 180 590 637] {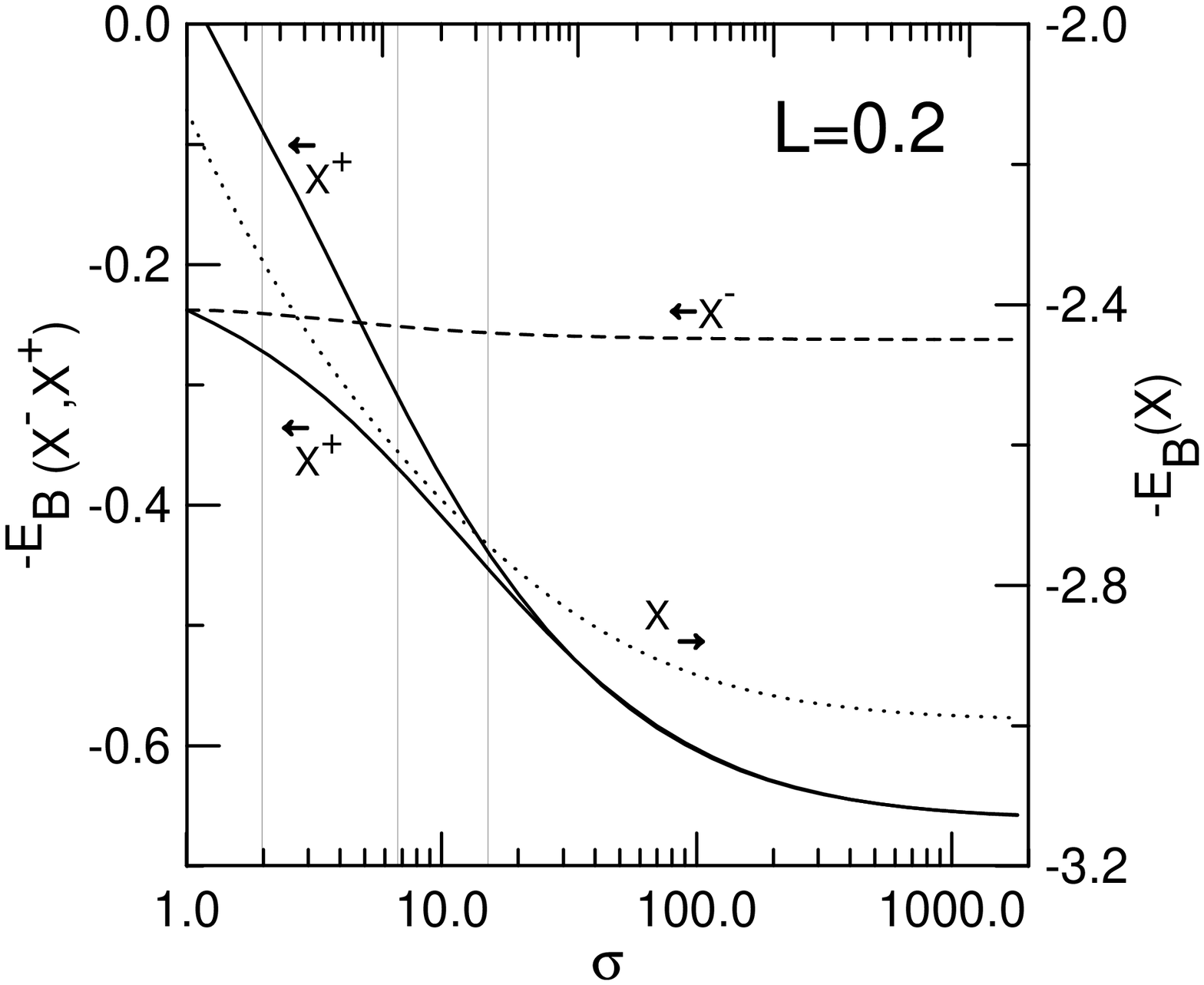}
\caption{Binding energies of the negative (dashed lines) and
positive trion (solid lines) states for $L=0.2$ as function of the
mass ratio $\sigma$. Higher solid curve corresponds to the $X^+$
state antisymmetric with respect to the interchange of electrons
and holes, ie. it is the first excited state. Thin vertical lines
show the values of $\sigma=1,1.98,6.72$ and 15.2. Dotted curve,
referred to the right axis, shows the exciton ground-state
eigenvalue. Energies and lengths are in donor units.
\label{psita}}
\end{figure}

\begin{figure}[htbp]
    \epsfxsize=60mm
                \epsfbox[30 93 580 605] {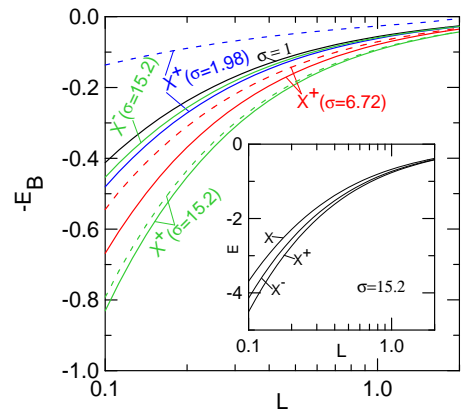}

\caption{[color online] Binding energies of the trions as
functions of the length of the lateral confinement. Lines for
$\sigma=1,1.98,6.72$ and 15.2 plotted with black, blue, red, and
green colors, respectively. For $\sigma=1$ binding energies of
$X^-$ and $X^+$ are equal. The dashed lines are the energies of
the excited $X^+$ states antisymmetric with respect to the
interchange of the holes ($X^-$ does not possesses a bound excited
state for $\sigma>1$). Lines for $X^-$ at $\sigma=1.98$ and 6.72
have been omitted for clarity - they are situated between the
$\sigma=1$ line and $\sigma=15.2$ line for $X^-$(see Fig.
\ref{psita}). Inset: energy eigenvalues for neutral exciton and
charged trions for $\sigma=15.2$. Energies and lengths are given
in donor units. }
\end{figure}

\subsection{Identical electron and hole lateral confinement}
For equal electron and hole lateral confinement ($l_e=l_h=L$) the
electron-electron, the hole-hole, and the electron-hole
interactions have the same form. The total interaction potential,
identical for both types of trions, is plotted in Fig. 1 (a) as
function of $z_{h1}$ and $z_{h2}$ for $L=1$. The regions of
positive (negative) potentials are plotted with red (blue) colors.
Zero of the interaction potential is marked with a dash-dotted
line. The interaction potential is minimal along the lines
$z_{h1}=0$ and $z_{h2}=0$ at which one of the two electrons and
the hole are in the same position (for $X^-$) or the position of
one of the holes coincides with the electron position (for $X^+$).
The potential is maximal along the diagonal $z_{h1}=z_{h2}$ at
which the two particles of the same charge are localized in the
same point along the wire length. Fig. 1(b) shows the
cross-sections of the interaction potential along the three
straight lines in Fig. 1(a) as function of the interelectron
($X^-$) or interhole ($X^+$) distance $z_{12}$. On the
antidiagonal the interaction has the form of a triangular
potential well [cf. solid line in Fig. 1(b)]. Moving along the
antidiagonal is equivalent to interchange the position of the two
particles of the same charge with fixed position of the third
particle of the opposite charge. Along the paths plotted with
dotted and dashed lines in Fig. 1(a), which are shifted below the
antidiagonal, the potential has the form of a double potential
well [see Fig. 1(b)] with a barrier near the diagonal resulting
from the repulsion of the equally charged particles. For large
$z_{12}$ the potential approaches $-3/z_{12}$ asymptotically which
is shown by the this solid line in Fig. 1(b).

Contour plots of the wave function of the negative and positive
trions calculated for different effective mass ratios are plotted
in Fig. 2 for $L=0.2$. For $l_e=l_h$ the negative and positive
trion relative Hamiltonians (7) and (8) differ only by the factor
standing in front of the mixed derivative ($1/\sigma$ for $X^-$
and 1 for $X^+$). The ground-state wave function for $\sigma=1$ is
the same for both trions [cf. Figs. 2(a,b)]. The first excited
state of $X^+$, which is antisymmetric with respect to the
interchange of the holes is unbound. Its wave function calculated
for the size of the computational box $100\times100$ (in donor
Bohr radius units) is plotted in Fig. 2(c). For the unbound state
no computational box is large enough (i.e. the wave function
vanishes only at the ends of the computational box). The wave
function is nonzero only near both axis. One of the holes stays at
the position of the electron and the other strives to be as far as
possible from the other two particles.

\begin{figure*}[htbp]
    \epsfxsize=100mm
                \epsfbox[83 490 538 750] {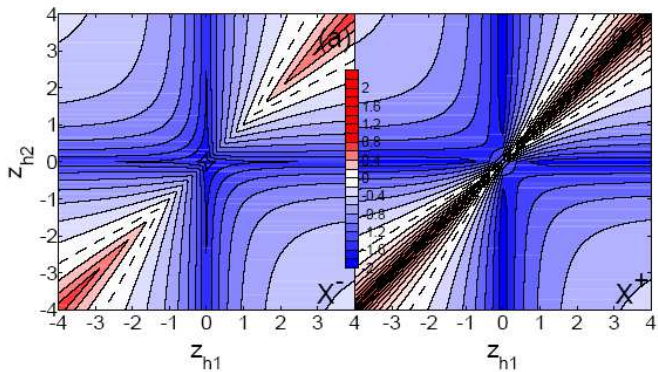}

\caption{[color online] Contour plot of the interaction potential
for (a) negative trion
$V=V^\mathrm{ef}(l_{e};z_{h1}-z_{h2})-V^\mathrm{ef}(l_{eh};z_{h1})-V^\mathrm{ef}(l_{eh};z_{h2})$
and (b) positive trion
$V=V^\mathrm{ef}(l_{h};z_{h1}-z_{h2})-V^\mathrm{ef}(l_{eh};z_{h1})-V^\mathrm{ef}(l_{eh};z_{h2})$
as function of the interparticle distances for the lateral
confinement lengths $l_e=1$ and $l_h=0.5$. Distances and energies
are given in donor units.}
\end{figure*}

Results for $\sigma=1.98$ plotted in Figs. 2(d-f) correspond to
CdTe material parameters ($m_h=0.19 m_0$, $m_e=0.096 m_0$) with
the donor units $a_d=5.4$ nm and $2R_d=27.6$ meV. The probability
density maximum for $\sigma=1$ is split into two extrema at the
antidiagonal of the plots [cf. Figs. 2(a-b)]. For $\sigma>1$ these
two extrema merge into a single one for $X^-$ [see Fig. 2(d)] and
for $X^+$ they become more distinctly separated [see Fig. 2(e)]
and the excited state for $X^+$ becomes bound [cf. Fig. 2(f)].

\begin{figure}[htbp]
    \epsfxsize=70mm
                \epsfbox[20 65 560 560] {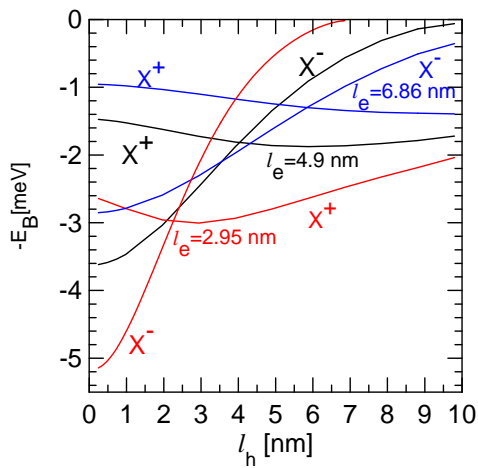}

\caption{[color online] Shifts of the trion recombination PL lines
with respect to the PL exciton line as function of the hole
confinement length ($l_h$) for GaAs. Different values of the
electron confinement length are plotted with different colors. }
\end{figure}

Figs. 2(g-i) and 2(j-l) for $\sigma=6.7$ and $\sigma=15.2$
correspond to GaAs ($m_h=0.45 m_0$, $m_e=0.067 m_0$, $2R_d=11.9$
meV, $a_d=9.8$ nm) and InAs ($m_h=0.41 m_0$, $m_e=0.027 m_0$,
 $2R_d=3.2$ meV and $a_d=29.7$ nm) material parameters,
respectively. Increasing $\sigma$ has an opposite effect on the
$X^-$ and $X^+$ wave functions. For the negative (positive) trion
the local minimum along the diagonal $z_{h1}=z_{h2}$ is less
(more) pronounced. The wave function evolution with $\sigma$ is
related to the tunnelling of the particles of the same charge via
the potential barrier presented in Fig. 1(b). The electrons in
$X^-$ with light effective masses tunnel easily through the
diagonal barrier due to the interelectron repulsion. On the other
hand the diagonal barrier is effectively much larger for the
heavy-mass holes which prevents its penetratation at large
$\sigma$ which leads to the appearance of the characteristic
maxima elongated along the diagonal in Figs. 2(e,h,k).

The correlation between the particles in the complexes is more
clearly visible in the pair correlation functions plotted in Fig.
3. The electron-hole correlation function is calculated as
$f_{eh}(z)=\int
d{z_{h1}}d{z_{h2}}|\chi(z_{h1},z_{h2})|^2\delta(z-z_{h1})$ and the
hole-hole (for $X^+$) , and electron-electron (for $X^-$) as
$f_{same}(z)=\int d{z_{h1}}d{z_{h2}}
|\chi(z_{h1},z_{h2})|^2\delta[z-(z_{h1}-z_{h2})]$. The Coulomb
hole in the hole-hole correlation in $X^+$ is much larger than for
electrons in $X^-$ but at the expense of slightly weaker
electron-hole localization.

The binding energies of the exciton and the trions for $L=0.2$ are
plotted as functions of $\sigma$ in Fig. 4. All the binding
energies are increasing functions of $\sigma$. In bulk the first
excited state of the positive trion is antisymmetric with respect
to the hole interchange,\cite{Frolov} possess the $P$ symmetry and
is bound for $\sigma>4.2$. The critical value of the mass ratio is
much smaller for quasi 1D confinement.\cite{Esser} Here, for
$L=0.2$ the excited $X^+$ state is bound for $\sigma>1.2$ (see
Fig. 3). For quasi 1D confinement the lowest excited state has the
$S$ symmetry with respect to the axis of the wire but is of odd
spatial parity, i.e., it is antisymmetric with respect to
simultaneous change of sign of all the $z$ coordinates (see Fig.
2).
 The ground-state of $X^+$ becomes degenerate with respect to the
symmetry of the wave function, i.e., the hole interchange, for
large $\sigma$ for which tunnelling through the diagonal potential
barrier (cf. Fig. 1) disappears. At large $\sigma$ also the
probability density of the excited $X^+$ level becomes identical
to the ground-state probability density (cf. Fig. 2). $X^-$ does
not possesses a bound excited state for $\sigma>1$.

The inset to Fig. 5 shows the ground-state energy for
$\sigma=15.2$ as function of the lateral confinement length. In
the $L=0$ limit the average interparticle distances decrease to
zero and the energies diverge to minus infinity. This is a
consequence of the Coulomb interaction singularity in one
dimension.\cite{Loudon}
 The
main part of Fig. 5 shows the shifts of the trion PL lines with
respect to the exciton line (calculated as the difference of the
eigenvalues presented in the inset) for different values of
$\sigma$. It turns out that the binding energies have a power law
dependence on $L$, i.e.  $L^{-q}$, for the $X^-$ and $X^+$ ground
state presented in this figure $q$ changes from $0.83$
($\sigma=1$) to $0.91$ ($X^+$ for $\sigma=15.2$).

\subsection{Effect of different electron and hole lateral confinement}
 Let us now consider the interaction potential for stronger hole
confinement. Figs. 6(a) and (b) show the interaction potentials
for $l_e=1$ as in Fig. 1 but for smaller $l_h=0.5$. For both the
negative [cf. Fig. 6(a)] and the positive trion [cf. Fig. 6(b)]
the potential minima at $z_{h1}=0$ and $z_{h2}=0$ become deeper
with respect to the $l_e=l_h$ case presented in Fig. 1. For $X^-$
the electron-electron interaction (the diagonal potential barrier)
is not affected by the change of $l_h$ [compare Fig. 1 and Fig.
6(a)]. On the other hand the hole-hole repulsive interaction for
$X^+$ is strongly increased.

\begin{figure}[htbp]
    \epsfxsize=70mm
                \epsfbox[14 190 575 715] {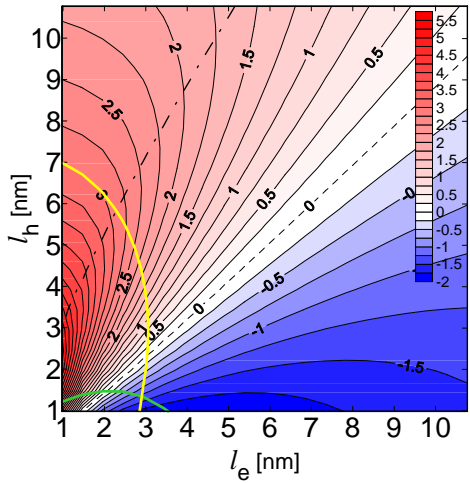}

\caption{[color online] Difference of the positive and negative
trion binding energies (in meV) as function of the electron and
hole confinement lengths for GaAs material parameters. Blue (red)
regions correspond to more stable negative (positive) trion. Above
the dashed-dotted line the negative trion is unbound. The green
line corresponds to $E_B(X^-)=4.2$ meV and the yellow line to
$E_B(X^+)=2.9$ meV.}
\end{figure}

\begin{figure}[htbp]
    \epsfxsize=60mm
                \epsfbox[30 40 560 575] {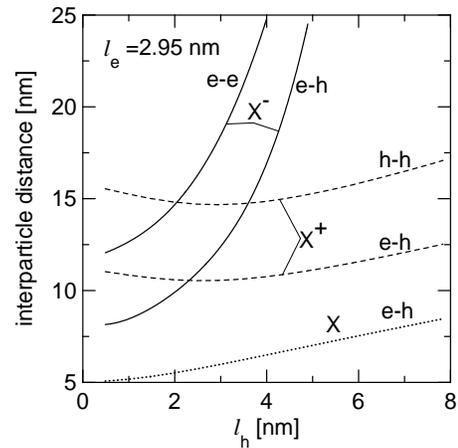}
\caption{Interparticle distances for the negative (solid lines)
and positive trions (dashed lines) and for the exciton (dotted
line) for GaAs material parameters and $l_e=2.95$ nm. (e-e),
(e-h), and (h-h) stand for the electron-electron, the
electron-hole and the hole-hole distance.}
\end{figure}

The effect of the hole localization on the trion binding energies
is plotted in Fig. 7 for GaAs material parameters and fixed values
of the electron lateral confinement. Consistently with the results
of section III.A for $l_e=l_h$ the positive trion is more stable
than the negative trion. A decrease of $l_h$ below the value of
$l_e$ results in the interchange of the $X^-$ and $X^+$ energy
lines.\cite{remark} This is due to the enhanced hole-hole
interaction shown in Fig. 6 (b). The negative trion binding energy
is a monotonous function of the hole confinement length, the
larger $l_h$ the smaller is the electron-hole interaction
stabilizing $X^-$. The situation is more complex for $X^+$, since
with increasing $l_h$ also the destabilizing hole-hole interaction
decreases. As a consequence the positive trion binding energy
possesses a maximum as function of $l_h$.

The difference of the positive and negative trion binding energies
is plotted in Fig. 8. Both the trions are equally stable for
$l_h=0.92 l_e-0.38$ nm. For $l_h$ larger (smaller) than $0.92
l_e-0.38$ nm $X^+$ is more (less) stable than $X^-$. The maximum
of the $X^+$ binding energy presented in Fig. 7 follows a path
that is nearly linear for $l_e>2$ nm and is approximately
parametrized by $l_h=1.62 l_e-1.98$ nm.
 For the points at the left of the dash-dotted line
the electron-hole attractive interaction stabilizing the trion is
so weak with respect to the electron-electron repulsive
interaction destabilizing the complex that the negative trion
stops to be bound (see also the line for $X^-$ at $l_e=2.95$ nm in
Fig. 7). The absence of the negative trion binding requires a
substantially weaker hole localization than the localization of
the electron which is rather impossible to obtain in the presently
produced quantum wires and would require the valence band offset
between the wire and the matrix to be much smaller than the
conduction band offset. Moreover, the present modelling based on
the assumption that the lateral wave functions are not affected by
the interaction is likely to fail since the hole wave function is
very likely to become more localized due to the attraction by
strongly confined electrons.

The fit of the calculated $X^-$ and $X^+$ binding energies to the
experimental data is obtained at the crossing of the green and
yellow lines, i.e., for $l_e=2.95$ nm and $l_h=1.3$ nm. The
obtained fit corresponds to realistic values which give a general
idea on the particle localization in the wire (the
measurements\cite{Otterburg} were performed on a V-groove
GaAs/AlGaAs quantum wire with a thickness of the GaAs crescent of
3 nm at the center). Obviously, a more realistic model is required
to extract details of the confinement from the experimental data.

The dependence of the size of the trion, i.e., the interparticle
distance as function of the hole confinement length, is shown in
Fig. 9 for $l_e=2.95$ nm. The electron-hole distance for the
trions and the exciton have been calculated as
$\sqrt{<z_{h1}^2>}$, and $\sqrt{<z_{eh}^2>}$, respectively. The
hole-hole distance for $X^+$ and the electron-electron distance
for $X^-$ are determined as $\sqrt{<(z_{h1}-z_{h2})^2>}$. The size
of the exciton increases as $l_h$ increases which is due to the
reduced value of the electron-hole interaction. Much stronger
dependence on $l_h$ is observed for $X^-$, which becomes unbound
for $l_h>7$ nm [cf. Fig. 7]. The dependence of the $X^+$ size on
the hole confinement is non-monotonous. The positively charged
complex has the smallest size near $l_h=3$ nm when it is the most
strongly bound [cf. Fig. 7]. For $l_h=l_e=2.95$ nm the order of
the interparticle distances in the two complexes is the same as in
two-dimensional quantum wells (compare Fig. 4 of Ref.
[\cite{RV}]). In spite of the fact that the probability of finding
both holes in $X^+$ in the same position is much smaller than for
electrons in $X^-$ (cf. Fig. 3) the longer tail of the
electron-electron correlation function results in a larger
electron-electron distance than hole-hole distance.

\begin{figure}[htbp]
    \epsfxsize=80mm
                \epsfbox[20 74 575 592] {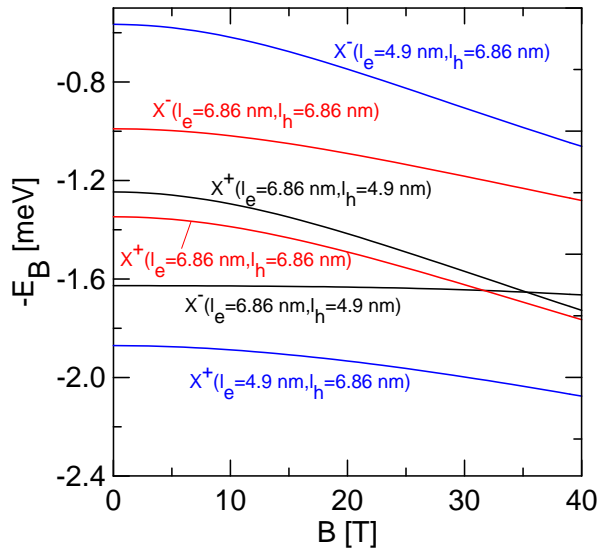}
\caption{ [color online] Magnetic field dependence of the trion
binding energies in GaAs for different values of the electron and
hole oscillator lengths. }
\end{figure}

\subsection{Magnetic field parallel to the wire}
In the present approach it is straightforward to include a
magnetic field oriented parallel to the axis of the wire. It
simply scales down the electron and hole oscillator lengths
according to the formula $ l(B)=(1/l^4(0)+1/l_c^4)^{-1/4}$, where
$l_c=\sqrt{2\hbar/eB}$ is the magnetic field length
($l_c=36.28/\sqrt{B}\mbox{ nm }\sqrt{\mbox{T}}$). Since $l(B)$ for
high magnetic fields decreases to $l_c$, the magnetic field tends
to equalize the electron and hole localization. The binding energy
of the trions can be obtained following paths of $(l_h(B),
l_e(B))$ on Fig. (8). The modification of the binding energies of
the trions by the magnetic field is presented in Fig. 10  for
different oscillator lengths. In a magnetic field of 40 T,
$l(B=0)=4.9$ nm is decreased to $l(B=40 \mathrm{T})= 4.4$ nm and
$l(B=0)=6.86$ nm to $l(B=40 \mathrm{T})= 5.19$ nm. For
$l_e=l_h=6.86$ nm (see the red curves in Fig. 10) the magnetic
field decreases the length of confinement exactly as in the case
presented in Fig. 5. In the more realistic case of stronger hole
confinement, i.e., for $l_e=6.86$ nm and $l_h=4.9$ nm (see the
black lines in Fig. 10) the magnetic field increases the binding
energy of the positive trion more strongly. This is because the
magnetic field more strongly affects the larger $l_e$ value than
the smaller hole localization length $l_h$ which increases the
electron-hole interaction more strongly than the repulsive
hole-hole potential. For $X^-$ the effect of the increased
electron-hole interaction is nearly cancelled by the increase of
the electron-electron potential value. Near 35 T a crossing of the
black lines is observed, which corresponds to passing from the
"blue" to the "red" region on the phase diagram of Fig. 8. This
crossing is qualitatively opposite to the one obtained for
two-dimensional quantum wells,\cite{RV} in which a small magnetic
field (around 1 T) increased the $X^-$ stability over the $X^+$
binding energy. In quantum wells the magnetic-field related
increase of the single-particle energy is smaller for electrons
and holes bound in the trion complex than for the lowest Landau
level in the final state of the free electron and hole after the
trion dissociation. The crossing of the binding energies observed
in quantum wires\cite{RV} is therefore at least partially due to
the stronger dependence of the electron lowest Landau level --
note an almost linear magnetic field dependence of the trion
binding energies in Fig. 2 of Ref. [\cite{RV}]. In the present
calculations the free electron and the free hole are strongly
localized in the plane perpendicular to the field and the
single-particle magnetic field effects cancel due to the
assumption of the frozen-lateral degrees of freedom, so that the
crossing is entirely due to the modified effective inter-particle
interactions.

 In the case of stronger electron confinement
($l_e=4.9$ nm, $l_h=6.86$ nm -- the blue lines in Fig. 10) the
situation is just opposite, the $X^-$ is less strongly bound and
the magnetic field acts more strongly on the negative trion
decreasing the binding energy difference according to the
mechanism described above. However, for the electron confinement
equal or stronger than the hole confinement, the magnetic field
does not lead to crossing of the trion energy lines and $X^+$ is
more stable for any magnetic field.

\section{Summary and Conclusions}
We studied the properties of the negative and positive trions in
quantum wires with strong lateral confinement using the
approximation of the lowest subband occupancy which allows for a
numerically exact solution of the multi-particle Schr\"odinger
equation. We investigated the relative stability of the positive
and negative trions with respect to the dissociation into an
exciton and a free carrier for different electron and hole
confinement. We found that the order of the negative and positive
trion PL lines is interchanged when the lateral confinement of the
hole is stronger than the one for the electron. In a GaAs quantum
wire with $l_e=5$ nm we predict that when $l_h$ is $20\%$  smaller
the positive and negative trion recombination lines interchange.
The change in the order is due to modification of the effective
interactions in the trion complexes. The present results provide
an explanation for the recently experimentally observed larger
stability of the negative trion in quantum wires.\cite{Otterburg}
We predict that for larger $X^-$ stability due to stronger hole
confinement the magnetic field oriented parallel to the axis will
tend to interchange the order of the $X^+$ and $X^-$ energy lines.
\vspace{0.5cm}

{\bf Acknowledgments} This paper was supported by the Flemish
Science Foundation (FWO-Vl), the Belgian Science Policy, the
University of Antwerpen (VIS and GOA), the EU-NoE SANDiE and by
the Polish Ministry of Scientific Research and Information
Technology within the solicited grant PBZ-MEN-MIN-008/P03/2003.
B.S. is supported by the EC Marie Curie IEF project
MEIF-CT-2004-500157. We are thankful for Daniel Oberli, Bart
Partoens and Yosip Sidor for helpful discussions.
\newline


\begin{thebibliography}{00}
\bibitem{Lampert} M.A. Lampert, Phys. Rev. Lett. {\bf 1}, 450
(1958).
\bibitem{Stebe2D} B. St\'eb\'e and A. Ainane, Superlattices
Microstruct. {\bf 5}, 545 (1989)
\bibitem{Riva2D} C. Riva and F.M. Peeters, Phys. Rev. B {\bf 61}, 13873 (2000).
\bibitem{Moradi} B. St\'eb\'e, A. Moradi, and F. Dujardin, Phys. Rev.
B {\bf 61}, 7231 (2000).
\bibitem{Glasberg} S. Glasberg, G. Finkelstein, H. Shtrikman, and
I. Bar-Joseph, Phys. Rev. B {\bf 59}, R10425 (1999).
\bibitem{Hawrylak} A. W\'ojs and P.
Hawrylak, Phys. Rev. B {\bf 51}, 10 880 (1995).
\bibitem{Whittaker} D.M. Whittaker and A.J. Shields, Phys. Rev. B
{\bf 56}, 15185 (1997).
\bibitem{Chapman} J.R. Chapman, N.F. Johnson, and V. Nikos
Nicopoulos, Phys. Rev. B {\bf 55}, R10221 (1997).
\bibitem{Esser} A. Esser, R. Zimmermann, and E. Runge, Phys. Stat.
Sol. (b) {\bf 227}, 317 (2001).
\bibitem{MCTs} T. Tsuchiya, Int. J. Mod. Phys. B {\bf 15}, 3985
(2001). \bibitem{Otterburg} T. Otterburg, D.Y. Oberli, M.-A.
Dupertois, N. Moret, E. Pelucchi, B. Dwir, K. Leifer, and E.
Kapon, (2004) cond-mat/0410050.
\bibitem{SM} B. St\'eb\'e and G. Munschy, Solid State Commun. {\bf 17}, 1051
(1975)
\bibitem{RV} C. Riva, F.M. Peeters and K. Varga, Phys. Rev. B {\bf
64}, 235301 (2001).
\bibitem{yusa} G. Yusa, H. Shtrikman, and I. Bar-Joseph, Phys.
Rev. Lett. {\bf 87}, 216402 (2001).
\bibitem{ClaraMC} A. V. Filinov, C. Riva, F. M. Peeters, Yu. E. Lozovik, and M.
Bonitz, Phys. Rev. B {\bf 70}, 035323 (2004).
\bibitem{Szafran1} B. Szafran, B. St\'eb\'e, J. Adamowski, and S.
Bednarek, J. Phys.: Condens. Mat. {\bf 12}, 2453 (2000).
\bibitem{Szafran2} B. Szafran, B.
St\'eb\'e, J. Adamowski, and S. Bednarek, Phys. Rev. B {\bf 66},
165331 (2002).
\bibitem{Belluci} D. Bellucci, F. Troiani, G. Goldoni, and E.
Molinari, Phys. Rev. B {\bf 70}, 205332 (2004).
\bibitem{Molinari} F. Rossi, G. Goldoni, and E. Molinari, Phys.
Rev. Lett. {\bf 78}, 3257 (1997).
\bibitem{Yosip} J. Maes, M. Hayne, Y. Sidor, B. Partoens, F. M. Peeters, Y.
Gonz\'alez, L. Gonz\'alez, D. Fuster, J. M. García, and V. V.
Moshchalkov, Phys. Rev. B {\bf 70}, 155311 (2004).
\bibitem{GaAs} A. Gustafsson, F. Reinhardt, G. Biasiol, and E. Kapon
 Appl. Phys. Lett. {\bf 67}, 3673 (1995).
\bibitem{InAsInP} L. Gonz\'alez, J. M. Garc\'ia, R. Garc\'ia, F. Briones, J. Mart\'inez-Pastor,
and C. Ballesteros, Appl. Phys. Lett. {\bf 76}, 1104 (2000).
\bibitem{CdTe} T.W. Kim, E.H. Lee, K.H. Lee, J.S. Kim, and H.L
Park, Appl. Phys. Lett. {\bf 83}, 4235 (2003).
\bibitem{Xia} J.-B. Xia and K.W. Cheah, Phys. Rev. B {\bf 55},
1596 (1997).
\bibitem{1D} S. Bednarek, B. Szafran, T.Chwiej, and J. Adamowski,
Phys. Rev. B {\bf 68}, 045328 (2003).
\bibitem{ktd} K.T. Davies, H. Flocard, S. Kreger, and M.S. Weiss,
Nucl. Phys. A {\bf 342}, 112 (1980).
\bibitem{Frolov} A.M. Frolov, and A. Yu. Yeremin, J. Phys. B: At.
Mol. Opt. Phys. {\bf 22}, 1263 (1989).
\bibitem{Loudon} R. Loudon, Am. J. Phys. {\bf 27} 649 (1959).
\bibitem{remark} The crossing point occurs exactly at $l_e=l_h$
only for $\sigma=1$.
\end{thebibliography}
\end{document}